\documentclass[conference]{IEEEtran}

\IEEEoverridecommandlockouts
\PassOptionsToPackage{hidelinks}{hyperref}
\usepackage{graphicx}
\usepackage{url}
\usepackage{amsmath}
\usepackage{amssymb}
\usepackage{booktabs}
\usepackage{multirow}
\usepackage{float}
\usepackage{placeins}
\usepackage{orcidlink}
\usepackage{cite}

\title{Adaptive Bridge: A Proxy-Based Decoupling Layer for Mitigating DDS Backpressure in ROS 2}

\author{
\IEEEauthorblockN{Kaushalraj Puwar \orcidlink{0009-0007-6354-3677}}
\IEEEauthorblockA{\textit{International Institute of Information Technology Bangalore}\\
Bangalore, India\\
Kaushalrajsinh.Puwar@iiitb.ac.in}
\and
\IEEEauthorblockN{B. Thangaraju \orcidlink{0000-0002-1960-6585}}
\IEEEauthorblockA{\textit{International Institute of Information Technology Bangalore}\\
Bangalore, India\\
B.Thangaraju@iiitb.ac.in}
}

\begin{document}

\maketitle

\begin{abstract}

In systems built on Robot Operating System 2 (ROS 2) and using Data Distribution Service (DDS), a single network-impaired or throttled subscriber on a RELIABLE topic can cause backpressure that degrades throughput and latency for all other subscribers, including safety-critical ones sharing the publisher, because the publisher's DDS writer can no longer accept new samples. We present Adaptive Bridge, a proxy-based layer that decouples critical subscribers from degraded or noncritical ones, thereby isolating the critical path through topic splitting and dynamic rate control. The proxy acts as a middleman and subscribes to the original topic and republishes the messages to two independent DDS writers: one RELIABLE writer for critical nodes and one BEST EFFORT writer for noncritical or degraded nodes, thus isolating the degraded nodes and safeguarding the publisher and critical nodes from backpressure. A probe-based classifier actively monitors subscriber health through sampling with hysteresis and adjusts subscriber rate limits in real time. We evaluate the system under a Gilbert-Elliott bursty wireless loss model using a reproducible Docker-based harness. The results show that using the Adaptive Bridge in our evaluation harness reduces the critical subscriber tail p95 latency from up to 15 s to 1.55 ms across all impairment severities while preserving the publisher's configured throughput.

\end{abstract}


\section{Introduction}

Consider a scenario in which a mobile robot publishes laser scans at 30 Hz, while a remote visualization node receives the data over Wi-Fi. Packet loss triggers Data Distribution Service (DDS) retransmissions, and the publisher stalls, leaving local collision avoidance without fresh data. The slow-subscriber backpressure coupling problem arises in Robot Operating System 2 (ROS~2) systems using the DDS middleware~\cite{ieee:ros2,sciangula2023,omg-dds}.

The coupling comes from the writer's handling of RELIABLE data. Samples remain in the writer history until all matched readers acknowledge them. A reader experiencing packet loss, limited bandwidth, or central processing unit (CPU) overload can leave samples unacknowledged. Once the history reaches its capacity, the publisher cannot write new samples. The writer blocks, the publisher drops messages, and healthy subscribers also see delayed delivery and reduced throughput~\cite{omg-dds,luo2025}. A degraded link affects every subscriber sharing that DDS writer.

Using BEST\_EFFORT for all subscribers removes this coupling, but it also gives up the delivery guarantees needed by safety-critical local subscribers. Quality of service (QoS) settings applied per subscriber do not change the shared writer's reliability policy, so a misconfigured or remote reader can still affect the pipeline. Differentiated Services (DiffServ) traffic prioritization has a different limitation: it operates on packets and does not use DDS semantics or subscriber criticality.

Adaptive Bridge places a middleware-level proxy between the publisher and its subscribers. The proxy subscribes to the original topic and sends the data through two independent DDS output writers. The critical path uses RELIABLE QoS for subscribers such as local collision avoidance; the noncritical path uses BEST\_EFFORT QoS, for subscribers such as remote visualization. This two-writer arrangement implements \emph{topic splitting} and interrupts the backpressure chain at the application layer without changes to DDS internals, ROS~2 Middleware Interface (RMW) implementations, or the publisher node. The proxy also runs a classifier that uses active round-trip time (RTT) probes and hysteresis to monitor subscriber health and adjust noncritical rate limits when impairment is detected.

The contributions of this work are:
\begin{enumerate}
  \item A proxy-based architecture for subscriber decoupling in ROS~2 that structurally isolates critical and noncritical data paths via independent DDS writers.
  \item An active-probe classifier with hysteresis for adaptive rate management on the noncritical path.
  \item A reproducible evaluation methodology using the Gilbert-Elliott bursty loss model on a Docker-based testbed with per-subscriber network impairment.
  \item Quantitative demonstration: publisher throughput preserved at 30 Hz (vs.\ a 29--36\% collapse in baseline) and critical subscriber tail latency reduced from up to 15 s to 1.55 ms at p95 across all impairment levels.
\end{enumerate}

\section{Related Work}

\subsection{DDS QoS and Performance Characterization}

RELIABLE and BEST\_EFFORT delivery in DDS determine how writers and readers match and how long samples remain in writer history~\cite{omg-dds,pardocastellote2003}. Sciangula et al. derive formal bounds for DDS data delivery under real-time constraints~\cite{sciangula2023}. Park et al. relate writer history depth to end-to-end delay in an analytical ROS~2 model~\cite{park2025}. Measurements across ROS~2 middlewares and deployment scenarios report throughput variability~\cite{chovet2025,maruyama2016}. The characterization stops at the coupling problem; no application-layer mechanism is proposed to separate the coupled paths.

\subsection{Adaptive QoS and Middleware Adaptation}

Runtime DDS frameworks adjust reliability and durability policies as conditions change~\cite{ingles2017}. Per-stream scheduling priorities have also been used in publish-subscribe communication~\cite{paikan2015}. Real-Time Publish-Subscribe Protocol (RTPS)/DDS-aware traffic handling has been studied for scalable wireless video streaming~\cite{almadani2013}. The changes occur at the publisher or transport level. Readers attached to the modified publisher still share one writer, so the policy change does not give each subscriber an independent data path or remove the shared reliability semantics.

\subsection{ROS~2 Real-Time and Backpressure Analysis}

Casini et al. examine response times in ROS~2 callback chains and analyze the impact of ROS 2 execution and scheduling behavior~\cite{casini2019}. Luo et al. quantify the communication-delay effect in a multi-subscriber ROS~2 setting; their analysis demonstrates the coupling effect and its impact on end-to-end latency~\cite{luo2025}. Middleware protocols for time-critical wireless transfer pursue similar throughput preservation goals, but operate below the ROS~2 application layer~\cite{peeck2021}. The analyses address the backpressure problem, and the protocol work addresses throughput at a lower layer. A deployable, noninvasive mitigation at the application layer is still absent.

\subsection{Contribution in Context}

Adaptive Bridge applies the mitigation at the ROS~2 application layer, without changes to DDS internals, RMW implementations, or publisher-side QoS reconfiguration. Its proxy places critical and noncritical subscribers on independent DDS writers, so the separation is in the writer structure rather than in a shared-writer QoS profile. The rate-control path uses active probes of per-subscriber network health and adjusts forwarding accordingly, instead of applying one static profile uniformly.

\section{System Design}

\subsection{Architecture Overview}

\begin{figure}[ht]
  \centering
  \includegraphics[width=0.75\columnwidth]{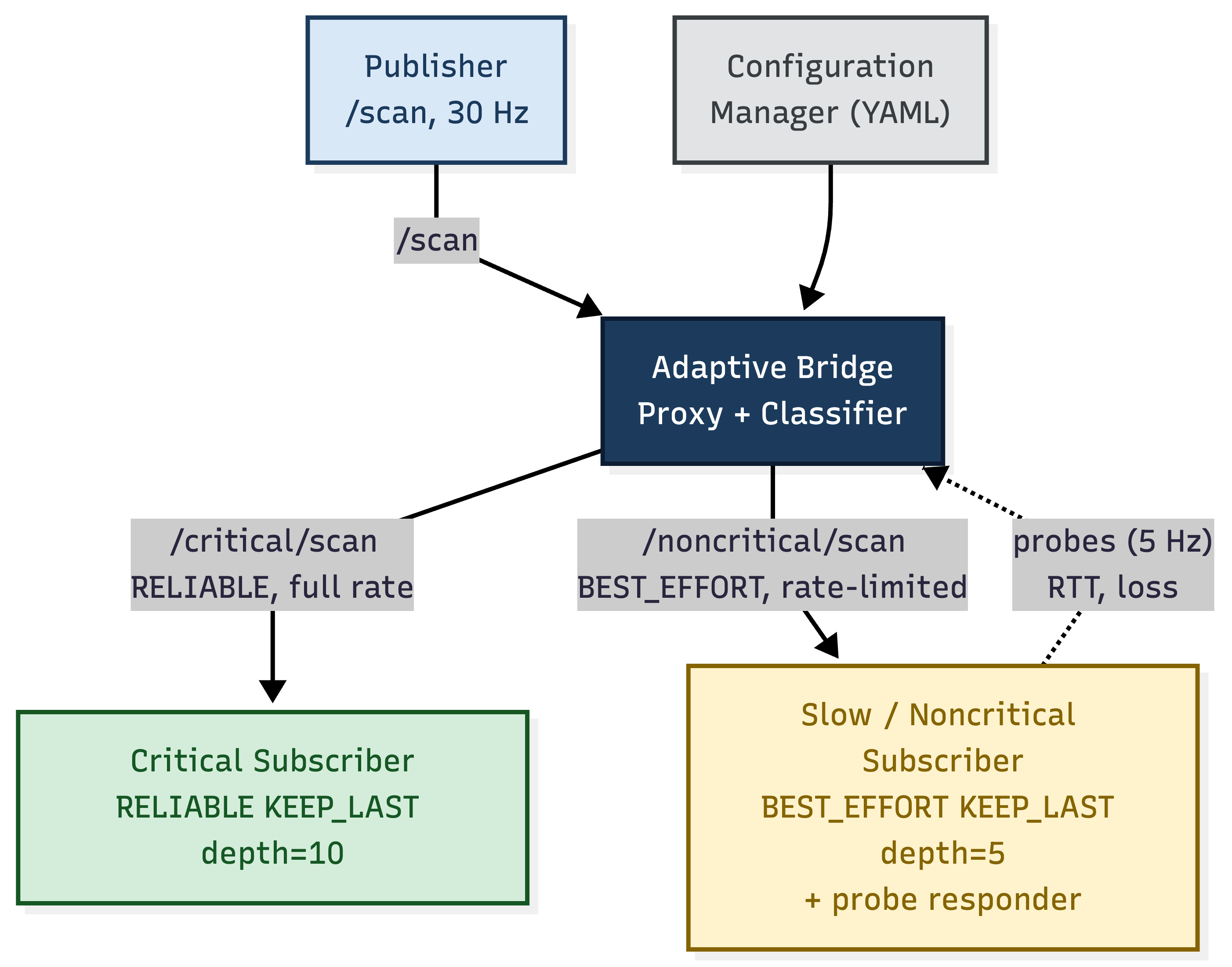}
  \caption{Adaptive Bridge architecture. The proxy receives the original topic and republishes it through two independent DDS writers. The classifier probes the slow subscriber and changes the noncritical forwarding rate.}
  \label{fig:architecture}
\end{figure}

Fig.~\ref{fig:architecture} shows the arrangement of Adaptive Bridge, which has four components. The Proxy Node receives input topics and republishes them on two independent output topics, one critical and one noncritical. Active probes from the Classifier Node provide subscriber-health measurements, and the Classifier Node publishes the resulting classification decisions. The Policy Engine uses those decisions to set rate limits and drop policies. The Configuration Manager reads the YAML configuration at startup, when all publishers are also created. Pre-creating them avoids discovery churn and runtime race conditions.

\subsection{Topic Splitting -- Isolation}

For an input such as \texttt{/scan}, the proxy publishes the received data through two independent DDS writers~\cite{omg-dds}. The critical writer uses RELIABLE QoS with KEEP\_LAST depth 10 and VOLATILE durability. The noncritical writer uses BEST\_EFFORT QoS with KEEP\_LAST depth 5.

Each writer matches its own readers. A slow reader on the BEST\_EFFORT writer cannot backpressure the RELIABLE writer because the DDS entities have separate history queues and acknowledgment schedules. This provides separation at the DDS endpoint level; it does not tune per-subscriber QoS. The critical readers are local and fast, so acknowledgments for the critical writer are received promptly. The publisher-to-proxy link does not accumulate unacknowledged samples, and the publisher is not blocked.

Park et al.'s analytical DDS latency model~\cite{park2025} shows that writer history queue length dominates end-to-end delivery delay. Splitting the writers gives each reader group its own queue. The critical queue is determined by the number of critical readers, typically one or two. The DDS data-delivery bounds from Sciangula et al.~\cite{sciangula2023} confirm that writer-level separation is the correct granularity for breaking backpressure.

Classification changes later affect only routing decisions and rate limits inside the proxy. No publisher is destroyed or recreated at runtime.

\subsection{Classifier -- Subscriber Health Monitoring}

\begin{figure}[ht]
  \centering
  \includegraphics[width=0.70\columnwidth]{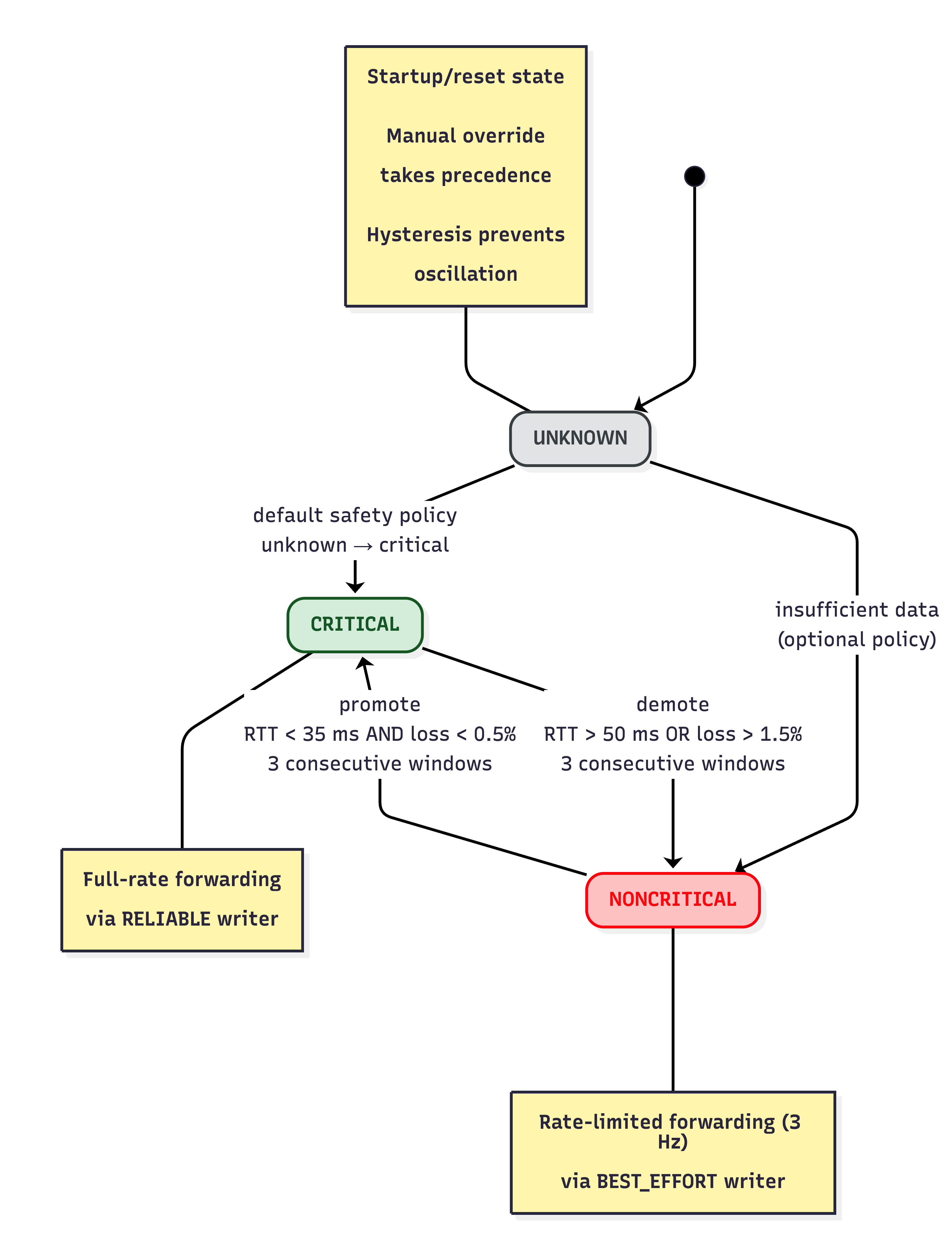}
  \caption{Classifier state machine. A state change requires threshold crossings over three consecutive classifier evaluations, while manual configuration overrides take precedence.}
  \label{fig:classifier}
\end{figure}

The classifier assigns each subscriber a link-quality state from active probes and maintains a per-subscriber state machine with hysteresis (Fig.~\ref{fig:classifier}).

\textbf{Probe mechanism.} Probe messages carry a sequence number. The proxy sends them at 5 Hz on a dedicated BEST\_EFFORT topic, and a responder at the subscriber returns the sequence number and timestamps. Over a sliding window of 50 samples (10 s), the classifier calculates mean round-trip time (RTT) and loss rate, where loss is the fraction of probes that receive no response.

\textbf{State machine.} A new subscriber starts in UNKNOWN. CRITICAL denotes a healthy subscriber whose data is forwarded at full rate through the critical writer. NONCRITICAL denotes a degraded subscriber whose data is rate-limited through the noncritical writer. UNKNOWN indicates insufficient probe data; it is treated as CRITICAL when \texttt{allow\_unknown\_state} is false.

Two threshold sets govern transitions and prevent oscillation. Demotion from CRITICAL to NONCRITICAL occurs when RTT exceeds 50 ms or loss exceeds 1.5\% for three consecutive classifier evaluations. Promotion back to CRITICAL requires RTT below 35 ms and loss below 0.5\% over the same number of evaluations. A reading between the promotion and demotion thresholds leaves the state unchanged and resets the evaluation counters. A manual override in the YAML configuration forces the specified state regardless of probe data.

The safety bias is explicit: misclassifying a critical subscriber as noncritical could drop safety-critical messages, whereas the reverse error only wastes bandwidth. The defaults and UNKNOWN handling reflect that bias.

\subsection{Policy Engine and Rate Limiting}

The Policy Engine selects forwarding behavior from the classifier state. In NORMAL mode, a CRITICAL subscriber receives all messages at the publisher's native rate through the critical RELIABLE writer. The noncritical path still forwards at the configured 10 Hz, providing a view of the stream without saturating bandwidth. A NONCRITICAL subscriber enters DEGRADED mode, where the noncritical BEST\_EFFORT writer is limited to 3 Hz by a token-bucket rate limiter.

The noncritical path also applies the following policies:
\begin{itemize}
  \item \textbf{Stale-drop:} messages older than 200 ms (configurable) are discarded before forwarding.
  \item \textbf{Queue-overflow protection:} all internal proxy queues are bounded; overflowing queues trigger backpressure on the noncritical path only.
  \item \textbf{Critical priority:} critical forwarding is always performed first; noncritical forwarding may be queued, throttled, or dropped but never blocks critical delivery.
\end{itemize}

\subsection{Safety Supervisor}

A global state machine watches internal queue occupancy, callback processing lag, and error counts to track proxy health. As overload approaches, the supervisor enters DEGRADED mode: noncritical forwarding is suspended entirely, while the critical path is preserved. Under extreme failure conditions, it
enters EMERGENCY mode, stops all forwarding, and publishes diagnostics. The supervisor prevents proxy overload from silently degrading critical-path delivery.

\section{Experimental Methodology}

\subsection{Evaluation Goals}

The evaluation tests three hypotheses. Hypothesis 1 (H1) concerns the baseline: bursty wireless loss causes DDS backpressure, which degrades publisher throughput and subscriber latency. Hypothesis 2 (H2) concerns the proxy: topic splitting through an intermediary proxy removes this degradation for critical subscribers by breaking the writer-level coupling. Hypothesis 3 (H3) concerns adaptation: compared with static topic splitting, adaptive classification provides an additional benefit by reducing noncritical bandwidth during impairment.

\subsection{Testbed Setup}

\begin{figure}[ht]
  \centering
  \includegraphics[width=0.8\columnwidth]{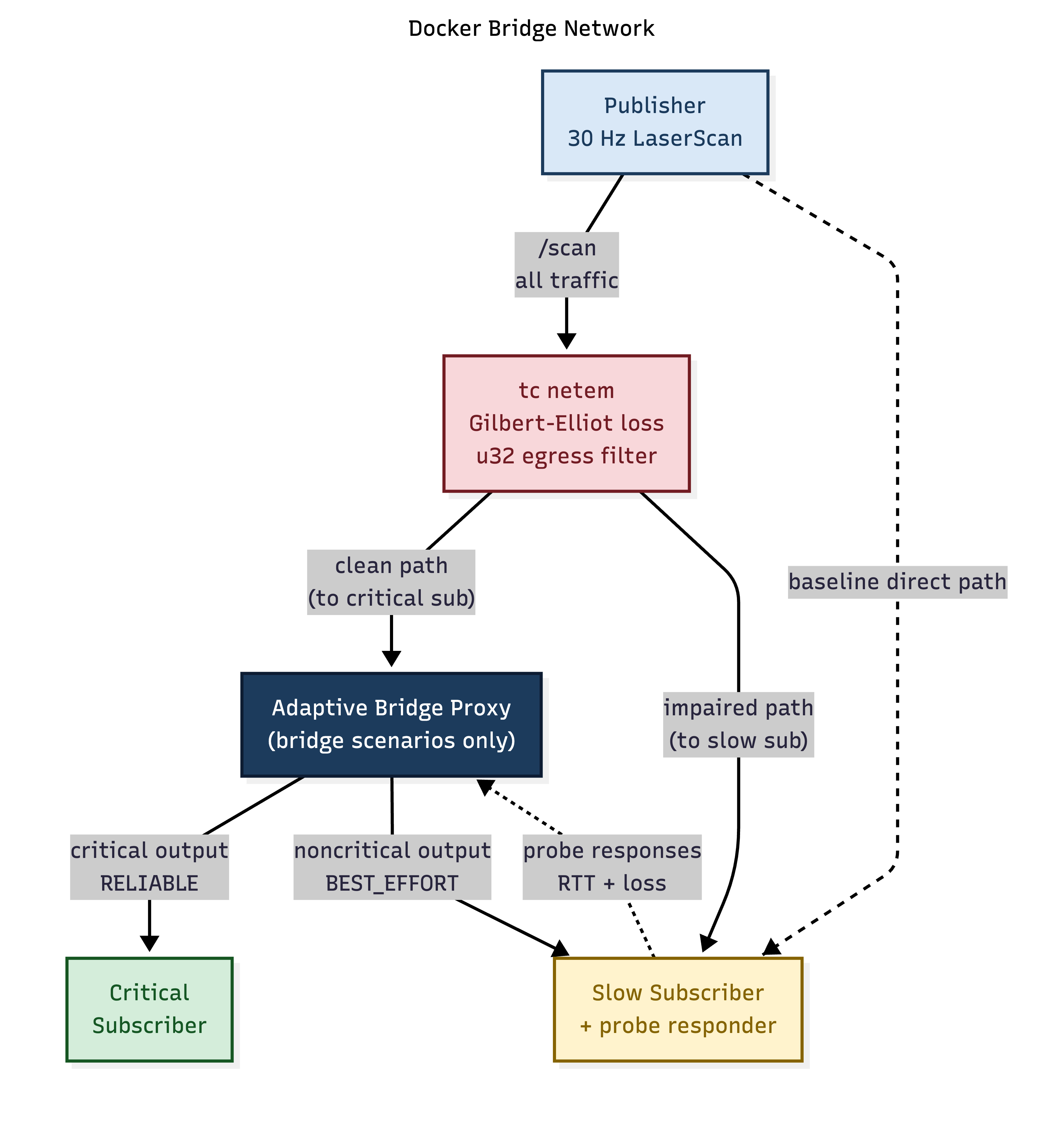}
  \caption{Experimental testbed topology. Four containers on a Docker bridge network. \texttt{tc netem} applies Gilbert-Elliott loss on publisher egress, filtered to the slow subscriber's Internet Protocol (IP) address only.}
  \label{fig:testbed}
\end{figure}

As shown in Fig.~\ref{fig:testbed}, the testbed uses four Docker containers on a bridge network. One container publishes \texttt{sensor\_msgs/LaserScan} messages at 30 Hz. In bridge experiments, a second container contains the proxy. The critical subscriber shares the local area network (LAN), while the slow subscriber represents a remote node subject to wireless impairment and runs a probe responder.

Fast DDS Extensible Markup Language (XML) profiles enforce User Datagram Protocol (UDP)-only transport by disabling shared memory (SHM) and specifying only the User Datagram Protocol version 4 (UDPv4) transport descriptor~\cite{chovet2025}. DDS traffic therefore crosses the container's network interface and is exposed to the \texttt{tc} rules. To reproduce the bounded-pool behavior observed in default Fast DDS configurations, the writer resource limit is set to \texttt{max\_samples=200}.

Linux \texttt{tc netem} applies the network impairment using the Gilbert-Elliott two-state Markov loss model~\cite{hemminger2005,jurgelionis2011,gilbert1960,elliott1963,wang1995}. The model switches between a Good state with low loss and a Bad state with high burst loss according to state-transition probabilities. The impairment is applied only to the slow-subscriber path; the critical subscriber path remains pristine. Probe responses receive an asymmetric return-path impairment of 20 $\pm$ 10 ms to model return-path network delay.

\subsection{Impairment Parameters}

The four evaluated impairment levels and their resulting loss characteristics are given in Table~\ref{tab:impairment}.

\begin{table}[h]
  \caption{Gilbert-Elliott Impairment Parameters}
  \label{tab:impairment}
  \centering
  \small
  \resizebox{\columnwidth}{!}{%
  \begin{tabular}{lccccc}
    \toprule
    Scenario & $p$ (G$\rightarrow$B) & $r$ (B$\rightarrow$G) & Good loss & Bad loss & Avg loss \\
    \midrule
    Clean     & --       & --   & --   & --   & 0\%    \\
    Mild      & 1\%      & 15\% & 0.5\%& 30\% & 2.3\%  \\
    Moderate  & 3\%      & 15\% & 0.5\%& 40\% & 7.1\%  \\
    Strong    & 2\%      & 10\% & 1.0\%& 50\% & 9.2\%  \\
    \bottomrule
  \end{tabular}%
  }
\end{table}

Impaired experiments run for 180 s and the clean experiment for 120 s, with up to approximately 5,400 latency samples at 30 Hz per run. The toggle experiment runs for 240 s, alternating impairment at 60 s intervals.

\subsection{Scenarios and Metrics}

The evaluation comprises ten experiments. Four are baseline runs, one for each impairment level, with no bridge. Five use the bridge with the classifier enabled. The remaining experiment is an ablation with the bridge present and the classifier disabled. The evaluation measures are critical-subscriber latency at p50 and p95, with p99 additionally reported for the cross-RMW comparison; publisher throughput rate and its standard deviation; noncritical-subscriber latency; and the classifier transition count.

The evaluation uses fixed, version-controlled configurations and impairment parameters, and the complete orchestration and analysis scripts are provided to enable reproducibility of the reported scenarios.

For each message, latency is the receive time minus the ROS message header timestamp; both publisher and subscriber use the ROS epoch clock. Publisher rate is computed from the number of publication attempts per 5-second window. To capture burst-driven throughput instability, rate standard deviation is computed over 5-second sliding windows.

\section{Results and Discussion}

\subsection{Baseline: Confirming Backpressure (H1)}

Table~\ref{tab:baseline} gives the baseline measurements. With no impairment, the publisher sustains 30.0 Hz with sub-millisecond latency. Under impairment, throughput collapses to 19.2--21.4 Hz, a 29--36\% drop, while critical-subscriber tail latency reaches 11.7--15.0 s. The 200-sample writer pool fills with unacknowledged samples as DDS backpressure develops, preventing the publisher from writing at its intended rate. The rate standard deviation of 8.5--8.7 Hz confirms burst-driven instability: slow-subscriber acknowledgments arrive as the pool fills and drains erratically. This simultaneous, measurable collapse in throughput and latency confirms H1 and results from DDS backpressure.

\begin{table}[h]
  \caption{Baseline Results (No Bridge)}
  \label{tab:baseline}
  \centering
  \small
  \resizebox{\columnwidth}{!}{%
  \begin{tabular}{lcccc}
    \toprule
    Scenario & Crit.\ p50 (ms) & Crit.\ p95 (ms) & Rate (Hz) & Rate std (Hz) \\
    \midrule
    Clean     & 0.65  & 0.95      & 30.0 & 0.0  \\
    Mild      & 0.97  & 15{,}013  & 21.4 & 8.5  \\
    Moderate  & 0.97  & 11{,}666  & 21.4 & 8.7  \\
    Strong    & 1.02  & 15{,}043  & 19.2 & 8.7  \\
    \bottomrule
  \end{tabular}%
  }
\end{table}

Median latency remains near 1 ms because messages experience negligible queuing before the 200-sample pool cap is reached. Backpressure therefore appears primarily in tail latency and throughput: slowed acknowledgments fill the pool until publisher skips reduce the measured rate.

\subsection{Bridge: Eliminating the Coupling (H2)}

\begin{table}[h]
  \caption{Bridge Results (Adaptive Bridge Enabled)}
  \label{tab:bridge}
  \centering
  \small
  \resizebox{\columnwidth}{!}{%
  \begin{tabular}{lcccc}
    \toprule
    Scenario & Crit.\ p50 (ms) & Crit.\ p95 (ms) & Rate (Hz) & Clf.\ trans.\ \\
    \midrule
    Clean     & 1.07 & 1.51  & 30.0 & 5  \\
    Mild      & 1.09 & 1.55  & 30.0 & 2  \\
    Moderate  & 1.10 & 1.55  & 30.0 & 2  \\
    Strong    & 1.09 & 1.55  & 30.0 & 2  \\
    Toggle    & 1.11 & 1.57  & 30.0 & 9  \\
    Ablation  & 1.09 & 1.57  & 30.0 & 0  \\
    \bottomrule
  \end{tabular}%
  }
\end{table}

Table~\ref{tab:bridge} shows the Adaptive Bridge measurements. Publisher throughput remains 30.0 Hz across bridge runs, with 5 s windowed rate standard deviation of 0.0 Hz at the reported one-decimal resolution. Critical-subscriber p95 latency does not exceed 1.57 ms across bridge runs, compared with 15.0 s in the baseline. The publisher-to-proxy link remains unblocked, which is consistent with the proxy draining its queue and the writer pool not accumulating.

The impaired runs show only two classifier transitions: after initial stabilization, sustained impairment causes a single demotion to NONCRITICAL, after which hysteresis prevents further changes. The clean run shows five transitions during startup and probe settling despite no network impairment; these affect only noncritical rate control and not the critical path.

Under strong impairment, publisher throughput rises from 19.2 Hz to 30.0 Hz and critical p95 latency falls from 15{,}043 ms to 1.55 ms, a factor of approximately 9,700. In clean mode, the bridge hop adds approximately 0.4 ms, increasing p50 latency from 0.65 ms to 1.07 ms, which is small relative to the baseline latency and throughput loss.

Topic splitting eliminates the backpressure coupling entirely: the slow subscriber and the critical subscriber use independent DDS writers. These measurements confirm H2.

\subsection{Cross-RMW Validation}

We repeated the evaluation matrix with Cyclone DDS (\texttt{rmw\_cyclonedds\_cpp}) under identical Gilbert-Elliott impairment conditions; Table IV reports the four bridge-enabled scenarios. This tests whether critical-path protection depends on Fast DDS; the harness switches RMW implementations with the \texttt{--rmw} flag without code changes.

For both RMWs, Table~\ref{tab:crossrmw} shows critical subscriber p99 latency below 2 ms and publisher throughput at 30.0 Hz across all impairment levels. The bridge's protection is RMW-portable across both implementations.

\begin{table}[h]
  \caption{Cross-RMW Validation: Bridge Scenarios}
  \label{tab:crossrmw}
  \centering
  \resizebox{\columnwidth}{!}{%
  \setlength{\tabcolsep}{3pt}
  \begin{tabular}{lcccc}
    \toprule
    Scenario & F-DDS crit p99 (ms) & C-DDS crit p99 (ms) & F-DDS rate (Hz) & C-DDS rate (Hz) \\
    \midrule
    bridge\_clean    & 1.7  & 1.9  & 30.0 & 30.0 \\
    bridge\_mild     & 1.8  & 1.9  & 30.0 & 30.0 \\
    bridge\_moderate & 1.8  & 1.6  & 30.0 & 30.0 \\
    bridge\_strong   & 1.8  & 1.6  & 30.0 & 30.0 \\
    \bottomrule
  \end{tabular}%
  }
\end{table}

The two RMWs handle \texttt{publish()} differently when backpressure occurs. Fast DDS uses non-blocking \texttt{publish()}. Once the \texttt{max\_samples=200} writer pool is full, the call returns an error while the timer continues to fire; messages then accumulate as a backlog, and baseline critical p99 latency reaches 14,000--17,000 ms. Cyclone DDS uses blocking \texttt{publish()}. When its 600 KiB writer history cache fills, \texttt{dds\_write()} blocks and prevents further message creation. No backlog forms, and critical p99 remains near 1 ms even under impairment. Fast DDS shows backpressure as an undelivered-message backlog, whereas Cyclone DDS shows it as reduced message generation. The topic-split architecture prevents the slow subscriber from causing backpressure in either implementation.

Cyclone DDS required one transport-level configuration adjustment, \texttt{AllowMulticast=spdp}, to force unicast data-plane traffic. This aligned its default multicast distribution with the per-subscriber \texttt{tc} filter, analogous to Fast DDS's explicit XML writer resource limits.

\subsection{Comparison Visualization}

\begin{figure}[h]
  \centering
  \includegraphics[width=0.55\columnwidth]{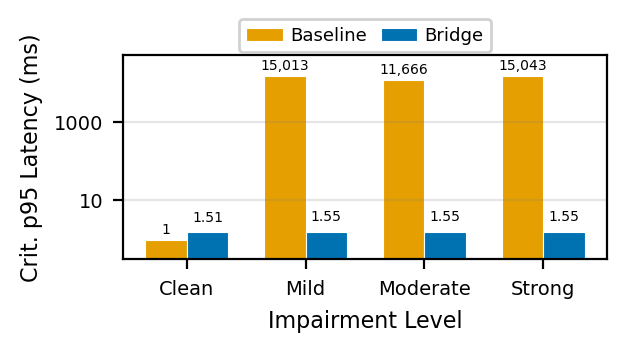} \\
  \includegraphics[width=0.55\columnwidth]{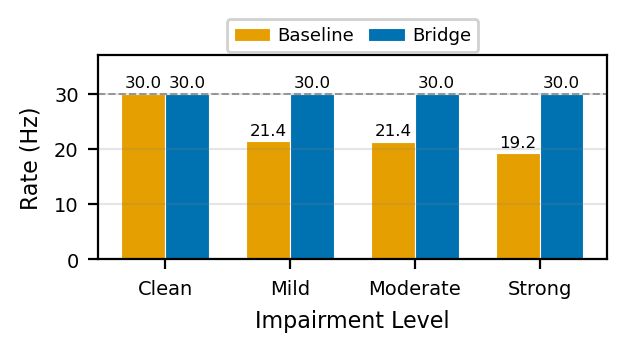}
  \caption{Critical subscriber p95 latency (top) and publisher throughput (bottom) across impairment levels. The bridge removes the two manifestations of backpressure. The dashed line indicates the publisher's 30 Hz target rate.}
  \label{fig:comparison}
\end{figure}

Fig.~\ref{fig:comparison} places critical subscriber p95 latency and publisher throughput side by side for the baseline and bridge at each impairment level.

\begin{figure}[h]
  \centering
  \includegraphics[width=0.55\columnwidth]{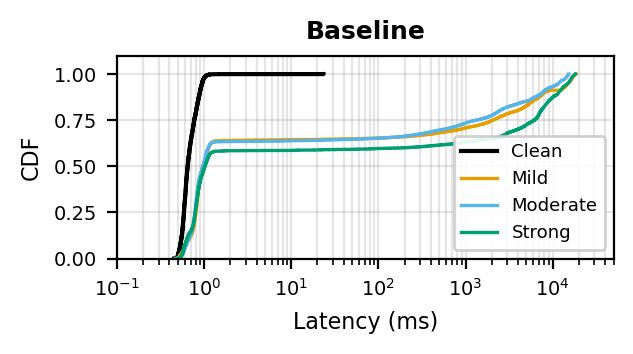}\\
  \includegraphics[width=0.55\columnwidth]{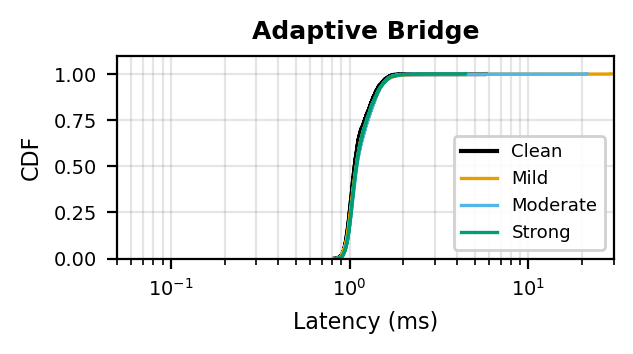}
  \caption{\textbf{Critical} subscriber latency CDFs across all impairment levels. The top panel is the baseline, where tail latency increases with impairment (clean below 1 ms; impaired tails exceed 18 s). The bottom panel is Adaptive Bridge, where all four CDFs overlap near 1--2 ms, showing no observable effect of impairment on critical-path delivery.}
  \label{fig:cdf}
\end{figure}

Fig.~\ref{fig:cdf} shows the critical subscriber latency cumulative distribution functions (CDFs) for both experimental conditions at all impairment levels.

\subsection{Ablation: Topic Splitting vs.\ Classification (H3)}

The ablation retained the topic split but disabled the classifier. Critical latency was p50=1.09 ms and p95=1.57 ms in the ablation, compared with p50=1.10 ms and p95=1.55 ms for bridge\_moderate. The two conditions are nearly identical, so the topic split alone provides the same critical latency benefit.

The classifier changes the noncritical path: with the classifier disabled, noncritical traffic continues at 10 Hz during impairment, while with it enabled, the rate falls to 3 Hz during degraded periods. On a bandwidth-constrained Wi-Fi link, the reduction from 10 Hz to 3 Hz during deep fades saves approximately 70\% of noncritical bandwidth and reduces proxy load and contention.

H3 is partially confirmed. The classifier adds adaptive control of noncritical bandwidth but does not improve critical-path protection beyond the static topic split. In the toggle experiment, impairment changed at 60 s intervals, producing 9 synchronized state transitions; detection and recovery occurred within approximately 20 s, consistent with the 10-second sliding window and three-evaluation hysteresis requirement.

\section{Limitations}

\textbf{Proxy overhead.} The bridge adds approximately 0.4 ms of latency. That overhead is negligible relative to the seconds of backpressure prevented by the bridge, but a proxy may become a bottleneck at high aggregate bandwidth. The evaluation focuses on network-induced backpressure and subscriber latency; CPU and memory overhead of the proxy were not independently measured and are therefore left for future evaluation.

\textbf{Reactive classification.} The classifier detects sustained impairments within approximately 20 s from a 50-sample sliding window; intermittent impairment shorter than the evaluation window may not cause a state change. Predictive classification based on network trends remains future work.

\textbf{RMW scope and message type.} The evaluation covers Fast DDS and Cyclone DDS under identical impairment conditions (see Section~V-C). Both implementations achieve equivalent critical-path protection, with sub-2 ms p99 latency, despite their fundamentally different \texttt{publish()} semantics. Additional RMWs, such as RTI Connext, and larger message
types, such as PointCloud2, remain to be evaluated.

\textbf{Controlled network model.} The Gilbert-Elliott parameters in Table I are used to produce controlled, bursty loss. These parameters are not claimed to replicate any specific real-world Wi-Fi deployment. Validation against empirically measured Wi-Fi traces remains future work.

\textbf{Single point of failure.} The Safety Supervisor monitors queue pressure, callback lag, and error counts and enters degraded or emergency modes under overload, but the proxy process remains a single point of failure. An external watchdog or redundant proxy configuration would improve resilience to process-level crashes.

\section{Conclusion}

The proxy-based topic split eliminates DDS backpressure coupling in mixed-criticality ROS 2 deployments. Across all impairment levels, the bridge preserves the 30 Hz publisher rate rather than the 29--36\% baseline collapse and reduces critical-subscriber p95 latency from 11.7--15.0 s to 1.55 ms. Topic splitting supplies the critical-latency benefit, while the adaptive classifier manages bandwidth on the noncritical path but does not provide additional critical-path protection.

The approach is deployable at the ROS~2 application layer without modifying DDS internals, RMW implementations, or the publisher node. Cross-RMW evaluation with Fast DDS and Cyclone DDS shows comparable critical-path protection under the evaluated impairments. Future work includes multi-proxy redundancy, evaluation across additional RMW implementations, and real-robot deployment validation. The complete implementation and configuration files are publicly available at the project repository~\cite{adaptive-bridge-github}, and the released ROS~2 package is indexed for the Jazzy distribution in the ROS Index~\cite{adaptive-bridge-ros-index}.

\end{document}